\documentclass[prl,twocolumn,preprintnumbers,amsfonts,a4paper]{revtex4-1}

\usepackage[sort&compress]{natbib}
\usepackage{graphicx} 
\usepackage{epstopdf}
\usepackage{dcolumn}	 
\usepackage{bm}		 
\usepackage{amssymb,amsmath}
\usepackage{latexsym}
\usepackage{color}
\usepackage[breaklinks,colorlinks,bookmarks=false,citecolor=blue,linkcolor=red,urlcolor=blue]{hyperref}

\graphicspath{{./}, {Figures/}}

\newcommand{\Eqref}[1]{Eq.~\eqref{#1}}
\newcommand{\Figref}[1]{Fig.~\ref{#1}}

\newcommand{\bea}{\begin{eqnarray}}
\newcommand{\eea}{\end{eqnarray}}

\newcommand{\be}{\begin{equation}}
\newcommand{\ee}{\end{equation}}

\newcommand{\e}{{\mathrm{e}}}

\newcommand{\TK}{T_{\mathrm{K}}}
\newcommand{\Tp}{T_{\mathrm{p}}}
\newcommand{\Mp}{M_{\mathrm{p}}}

\newcommand{\HH}{{\cal{H}}}
\newcommand{\TT}{{\cal{T}}}

\newcommand{\Eaverage}[1]{\langle{#1}\rangle}

\begin{document}

\title{A discrete energy space induced fermion parity breaking fixed point of the Kondo model}

\author{Peter Schmitteckert}
\email{Peter.Schmitteckert@physik.uni-wuerzburg.de}
\affiliation{Lehrstuhl f{\"u}r Theoretische Physik I, Physikalisches Institut, Am Hubland, Universit{\"a}t  W{\"u}rzburg, 97074 W{\"u}rzburg, Germany}

\date{\today}

\begin{abstract}
In this work we combine the well established  Kondo problem with the more speculative field
of a discrete space time. We show that a discrete energy space induces a flow towards a new fixed point
by breaking the conservation of charge and spin  and  lifting the fermion parity. This parity lifting fixed point appears on
a scale set by the discretization of energy space. In contrast to the Planck scale the associated energy scale
is at very low energy scales, possibly given by the inverse size of the universe.
\end{abstract}



\maketitle

\section{Introduction}

One of the most striking features of quantum mechanics is the non-locality of entanglement,
which has recently been verified by several independent experiments  \cite{Roger:PRL82,Gisin:Nature98,Zeilinger:PRL98,Aspect:Nature99,Wineland:Nature01}
testing Bell's inequality \cite{Bell:P64},
including impressive bounds \cite{Zbinden:Nature08} on the minimal speed. Taking this concept to its ultimate limit 
every particle should be able to explore the complete universe, even currently unknown parts of the universe possibly containing new physics. 
At least we can not rule out such an entanglement.
In return  we should treat every particle as a particle in a box, namely the universe,
which leads us to the conclusion, that there is a discrete space--time at very low energies.
Such an hypothesis opens the possibility for rather unconventional, speculative scenarios,
e.g. a deterministic description of the world which does not violate Bell's inequalities \cite{tHooft:X12,tHooft:X16}.
A discrete world is also seen as a possibility to overcome the fundamental difference between general relativity
and quantum mechanics, leading to concepts like quantum foam \cite{Wheeler:PR55} or non-commutative geometry \cite{Witten:NPB86,ConnesLott:NPPS89}.
There one actually assumes a minimal length scale, see also \cite{Hossenfelder:LRR13}, corresponding to a high energy scale set by the
Planck scale.
In contrast we are studying the case where the world is discrete at low energy scales.
Assuming a size of universe of $L=9.1\,10^{10}$ light years \cite{Itzhak:09} quantum mechanics implies an energy scale $E_{\mathrm{p}}$
for our particle in the universe system of
\begin{equation}
	E_{\mathrm{p}} \,=\, \frac{ \hbar c}{L} \,\approx\, 3.6\,10^{-35} \,eV \,,
\end{equation}
with $c$ the speed of light, and $\hbar$ the reduced Planck constant. While this scale is far below any current experimental resolution
there might exist emergent phenomena influencing physics to much higher, experimentally accessible, scales.  
Alternatively one may look at quantities where single particle effects can lead to a drastic change of observables.
Since a discretized world naturally breaks all sorts of symmetries the search for an  unusual breaking of a symmetry is an interesting
candidate for such an observable.
Here we study the fermion parity ${P} = N \mod 2$, with $N$ the number of fermions in the universe, which has the interesting property
that it is sufficient to break the fermion parity for a single fermion to destroy the symmetry completely, despite $N$ being very large.
Note, that the fermion parity is either zero or one. We can't just perturbativly change it a little bit without destroying the symmetry,
So, if somewhere, even in the far distant parts of the universe, an unknown mechanism breaks the fermion parity, we might be able to
observe it in the lab. In the following we demonstrate that a discretized world itself may already lead to such a breaking of the fermion parity.

As a system to study the conservation of the fermion parity we look at the Kondo model.
The Kondo model and the associated Kondo resonance can be seen as the prime examples
for correlated quantum systems. The Kondo problem itself can be traced back to the experiments
by de Haas  and van den Berg \cite{deHaas:P36} in the early 1930s which displayed an increase in resistivity
of noble metals like gold or silver. It took 30 years until Kondo \cite{Kondo:PTP64,Kondo:SSP69} could relate the increase of
resistivity to dynamical scattering at magnetic impurities. But it was only more than 30 years later
that Wilson could provide a rigorous solution of the Kondo model based on his
numerical renormalization group (NRG) technique \cite{Wilson:RMP75,Wilson:Adv75,Hewson:93,CostiPruschke:RMP08}.
A few years later Andrei \cite{Andrei:PRL80} and Vigman \cite{Wiegman:JETP80} could verify the numerical solution
of Wilson by a Bethe ansatz solution.
Besides its importance in describing magnetic impurities in metals it is also important in understanding
the transport properties of quantum dots \cite{GoldhaberGordon:PRL98} and often appears as effective model
in understanding correlated quantum systems, e.g. the dynamical mean field theory \cite{MetznerVollhardt:PRL89,Georges:RMP96}. 
For an overview see \cite{CostiPruschke:RMP08}.

The striking feature of the Kondo model, where a spin 1/2 is coupled to a conduction band via exchange coupling,
is that the system consist of two regimes: the weak coupling (WC) and the strong coupling (SC) regime 
and the NRG allows us to study the crossover between these regimes.
In the weak coupling the system looks like a local moment, i.e. the spin impurity, and a separated conduction band
which are perturbativly coupled, while in the SC regime the local moment is screened by the conduction band forming a singlet.
In summary the Kondo model and its solution via the NRG technique are by now  well studied and the NRG is the reference
method of choice for equilibrium quantum impurity problems,

The last ingredient missing in our study is a descretized representation of the universe.
Here we assume that the energy discretization is a fundamental property and not just a consequence of the particle in the universe problem,
e.g. commutation relations are only valid up to ${\cal{O}(E_{\mathrm p}})$.
To this end we point out  that typically all simulations on todays computers are performed within a discrete description.
For instance the IEEE 754 description \cite{IEEE754} of a {\tt binary64} ({\tt double} precision) number 
consists of 64 bits, 53 \footnote{Since the leading binary digit is always a 1 it
is not stored, so it only takes 52 bits in memory.} for the mantissa,
one sign bit, and 11 bits for the exponent  leading to 15 to 17 significant digits representing the resolution of the world.
While it is not surprising that the limited precision in representing real numbers  may eventually change the results
of the iterative RG procedure, it is surprising that instead of obtaining a numerical mess the system flows to a new fixed point
where fermion parity is lifted. 

\section{Kondo model}
The Kondo model describes a local impurity coupled to a conduction band. Following section {\tt VII} and {\tt VIII} of \cite{Wilson:RMP75} the 
conduction band is transformed into spherical harmonics around the impurity and only the $s$-wave contribution is kept.
The remaining model of a spin impurity coupled to a half infinite chain is then discretized on a logarithmic scale.
The system is then tridiagonalized leading to the following form:
\begin{align}
	\HH_M &= J \vec{\hat{S}} \vec{\hat{s}}_1 \,+\, \sum_{n=2}^{M}  t_{n-1}  \sum_\sigma \hat{c}^\dagger_{n,\sigma}  \hat{c}^{}_{n-1,\sigma} \,+\, \text{h.c.} \,. \label{eq:KondoModel}
\end{align}

Here we follow the usual convention of $\vec{\hat{S}}$ being the $SU(2)$ spin operator of the impurity,
 $\vec{\hat{s}}_1$ is the spin operator of the first conduction band site, $\hat{c}^{}_{n,\sigma}$ ($\hat{c}^\dagger_{n,\sigma}$)
is the annihilation (creation) operator for a conduction band fermion with spin $\sigma$ in energy shell  $n$. 
$J$ denotes the Heisenberg exchange coupling and 
\begin{align}
	t_n = t \Lambda^{(n-1)/2} \label{eq:tn}
\end{align}
is taken in its most simplistic form ignoring any corrections stemming from the original band structure
as we are interested in very low energy physics only. For a justification for this Hamiltonian we refer to 
excellent articles by Wilson \cite{Wilson:RMP75,Wilson:Adv75}.  
However, for this work it is sufficient to know that $\HH$ \eqref{eq:KondoModel} describes a single spin
coupled to a 1D like tight binding chain where the $n$-th site represents the physics at
energy scale $t\, \Lambda^{(n-1)/2}$.

\section{Kondo RG flow}
\begin{figure}[!t]
 \includegraphics[width=0.49\textwidth]{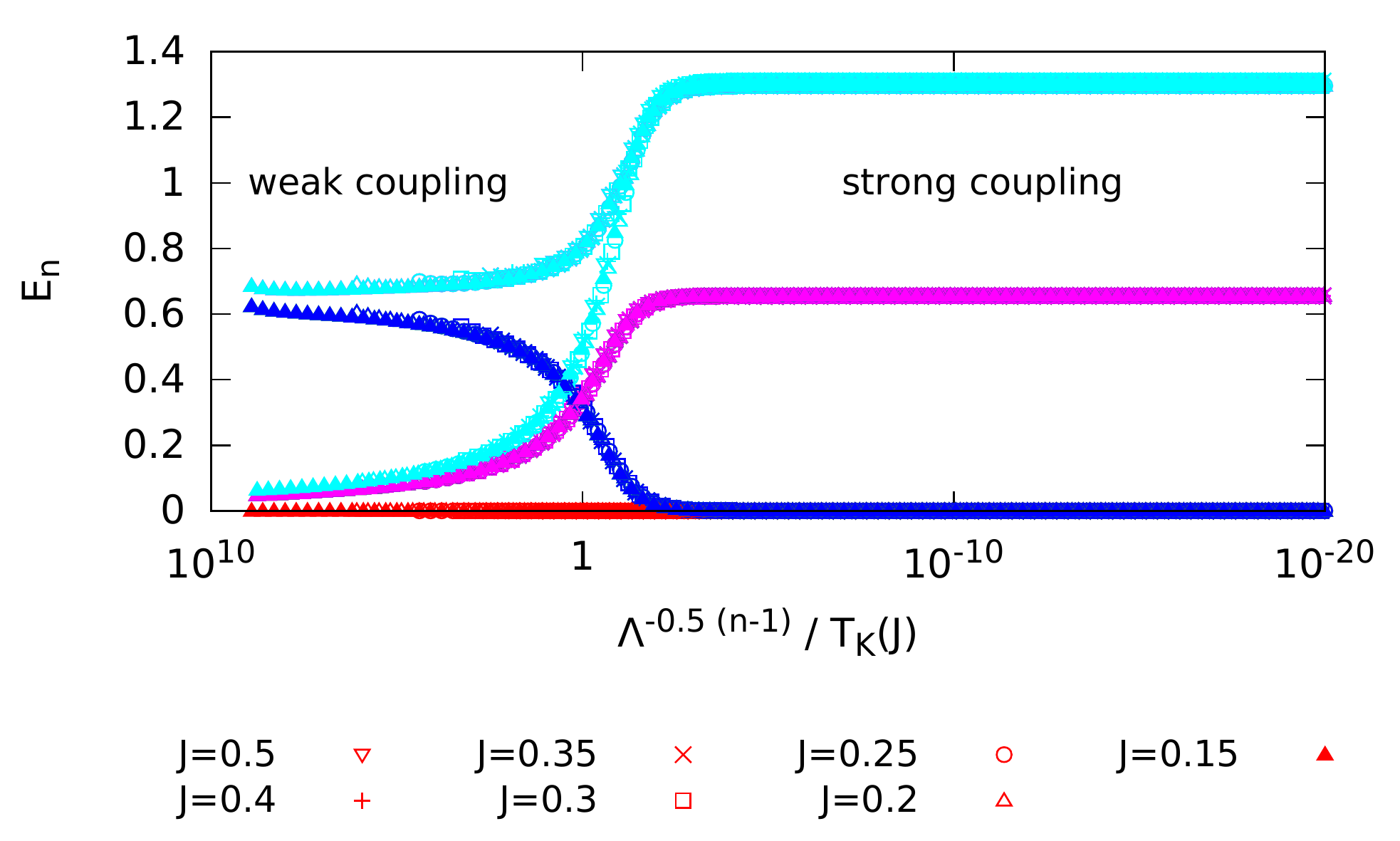} 
 \caption{\label{fig:RGFlowKondo} The RG flow of the low energy spectrum for a Kondo model with coupling
 $J=0.15$, $0.2$, $0.25$, $0.3$, $0.35$, $0.4$, and $0.5$, $\Lambda=2.0$, and $m=2000$.
	The RG scale $t_n=\Lambda^{-(n-1)/2}$ is expressed in units of the Kondo temperature $\TK(J)$.
   }
 \end{figure}
The important feature of the Wilson NRG is the existence of a fixed point $\widetilde{\HH}^*$ in the low energy regime of the 
(half-) group iteration $\TT$
\begin{align}
	\widetilde{\HH}_M                 &= \Lambda^{M/2} \HH_M \label{eq:KondoTilde} \\
	\TT\left[\widetilde{\HH}_M \right] &= \tilde{\HH}_{M+1} \\
	\TT^2\left[\widetilde{\HH}^* \right] &= \widetilde{\HH}^* \label{eqFixPoint}
\end{align}
The reason for rescaling $\HH_M$ in \eqref{eq:KondoTilde} consists in the fact that we are interested in the low energy scales.
Naturally, by adding a new energy scale in $\HH$ one obtains levels which are a factor of $\Lambda^{-1/2}$ smaller.
So one could never obtain a fixed point. Instead one blows up the Hamiltonian precisely by this factor.
In return by adding a new layer / site one first rescales the current Hamiltonian by  $\Lambda^{1/2}$ and couples the 
new site with a hybridization of $t$. Of course, when we are interested in the associated physical energy scale
we have to undo the rescaling of \eqref{eq:KondoTilde}.
Wilson \cite{Wilson:RMP75} has shown that the energy spectrum of the rescaled Hamiltonian can be used to 
classify regimes of the Hamiltonian of interest.
It turns out \cite{Wilson:RMP75} that for the Kondo model one obtains two regimes: weak (WC) and strong coupling (SC), 
see Fig.~\ref{fig:RGFlowKondo}.
For a small number of lead sites $M$, corresponding to high energy scales, the systems looks like a free spin that is
weakly coupled to $M$ conduction bath sites, while at low energy scales, i.e. large $M$, the impurity spin forms
a singled with one conduction band electron, which acts as a weak perturbation to effectively $M-1$ conduction band sites.
The striking feature of the Kondo model is the appearances of a scale $\TK$
\begin{equation}
	\TK = D \sqrt{J/D} \,\e^{-D/J} \label{eq:TK} \,
\end{equation}
with the $D=4t$ the band width for the Hamiltonian \Eqref{eq:KondoModel} and $t$ the band hopping element \Eqref{eq:tn}.

In \Figref{fig:RGFlowKondo} we show the flow of the five lowest excitation energies vs.\ the energy scale 
$T_M = \Lambda^{(M-1)/2}$ in units of $\TK$. In this computation the particle number and the $S^z$ component of the 
total spin where explicitly conserved by working with a block matrix representation of the Hamiltonian.
Note that within the NRG scheme \cite{Wilson:RMP75,Wilson:Adv75} the ground state energy $E_0$ is shifted to zero.
The collapse of the data points, when rescaled with the Kondo temperature, demonstrates the single parameter scaling 
of the Kondo problem: physics is governed by the scale $\TK$ \Eqref{eq:TK} which is exponential in $1/J$.
In addition, the NRG procedure is extremely stable, even down to scales which are more than twenty orders in magnitude smaller
compared to the Kondo scale, which itself is already a low energy scale.
\begin{figure}[!t]
 \includegraphics[width=0.49\textwidth]{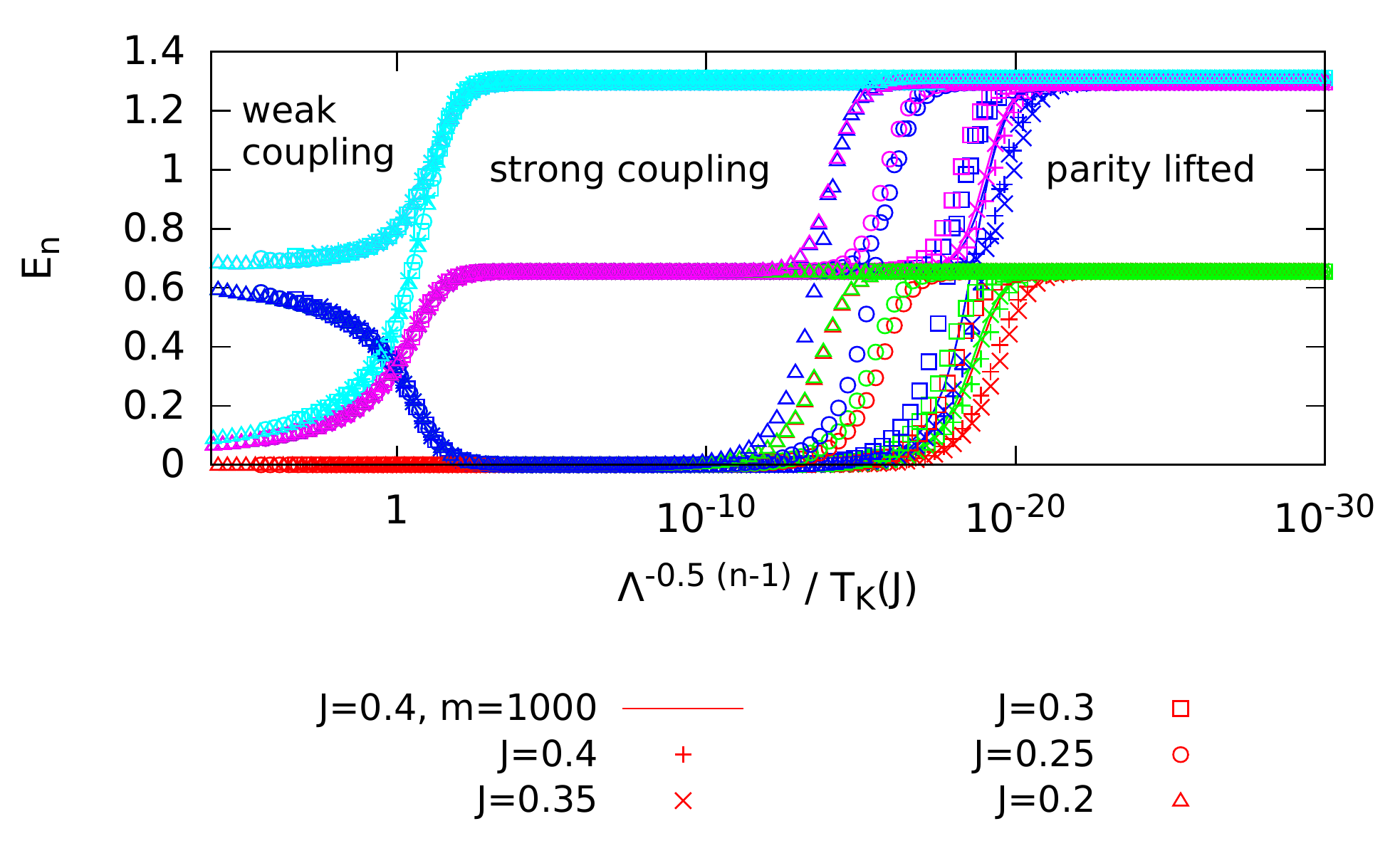} 
 \caption{\label{fig:RGFlowKondoNC} The RG flow of the low energy spectrum for a Kondo model with couplings
 $J=0.2$, $0.25$, $0.3$, $0.35$, and $0.4$, $\Lambda=2.0$, and $m=2000$, where the $N$ and $S^z$ conservation is not explicitly enforced. 
 In addition the lines show the results for $J=0.4$ and $m=1000$ states kept.
 The RG scale $t_M$ is expressed in units of the Kondo temperature $\TK(J)$, see \Eqref{eq:TK}.}
 \end{figure}

\section{Fermion parity lifting fixed point}
The situation changes if we do not enforce the symmetries, like spin and charge conservation, in the NRG code.
In \Figref{fig:RGFlowKondoNC} we show the the NRG flow in simulations without enforcing any symmetries.
For not too small energy scales we obtain precisely the same results as with enforced symmetries. Of course,
deviating results would imply an error in the code.
However, in contrast to a standard NRG simulation as in \Figref{fig:RGFlowKondo}, we now witness the flow towards a  new fixed point
at extremely small energy scales which clearly is not related to the Kondo scale $\TK$.
In \Figref{fig:RGFlowKondoNC} we have added the results for $J=0.4$ and $m=1000$  instead of $m=2000$ states as used for the results
represented by symbols. While the results still change with $m$, the overall picture does not change.
\begin{figure}[t]
 \includegraphics[width=0.49\textwidth]{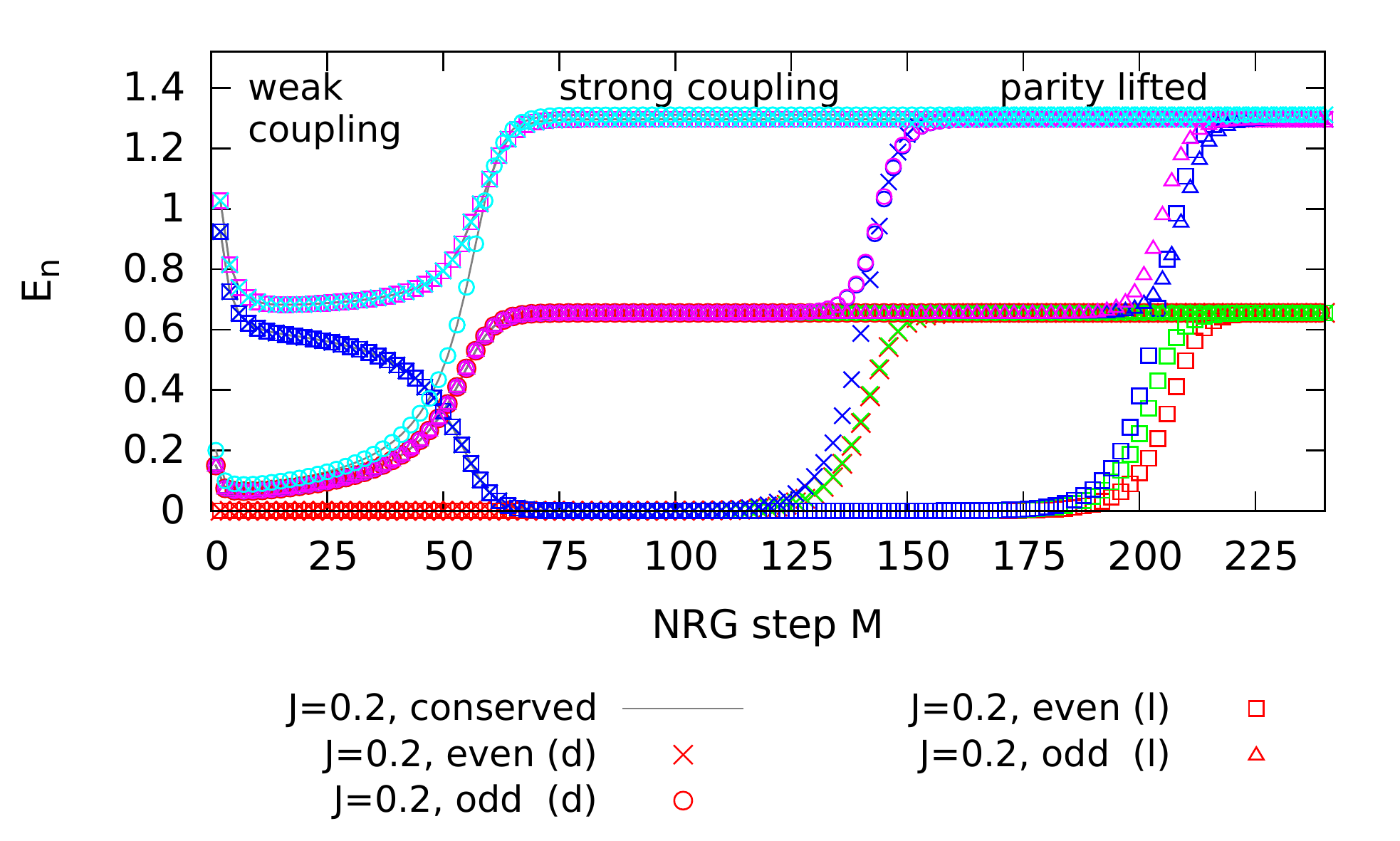} 
 \caption{\label{fig:RGFlowKondoParity} The RG flow of the low energy spectrum for a Kondo model with coupling
 $J=0.2$, $\Lambda=2.0$, and $m=2000$  vs.\ the NRG step (lead size) $M$,  where the $N$ and $S^z$ conservation is not explicitly enforced. 
 The crosses and the squares are the even $M$ sector, and the circles and triangles 
 are the results in the odd $M$ sector. The crosses and the circles are calculated within
 {\tt double} precision, the squares and the triangles are calculated using {\tt long double} precision.
 The line shows the results for a calculation where particle number
 and the total spin $S^z$ component is explicitly conserved.}
\end{figure}

In order to investigate this new  fixed point we show in \Figref{fig:RGFlowKondoParity} the results for $J=0.2$ vs.\ the NRG step,
i.e. the lead size, $M$. As mentioned before, a property of the NRG solution is, that there are always two fixed points for a given
coupling. There reason is actually quite simple. The uncoupled lead has a unique ground state at half filling only for an
even number of sites, for an odd number of sites there is a twofold degeneracy corresponding to $\Eaverage{\hat{S}^z}=\pm 1/2$ as
the number of fermions is odd. 
In the limit of $J\rightarrow 0$ we have an $M$-site chain perturbed by a local moment. In the $J\rightarrow \infty$ limit,
the spin impurity and a conduction band electron form a singlet. Therefore there is one fermion less in the conduction band
and the odd-even effect is reversed.
Indeed, the cross-over at the Kondo scale $\TK$ exchanges the odd and even sectors.
That is also the reason why there are only fixed points for ${\cal T}^2$ and not for ${\cal T}$ in \Eqref{eqFixPoint}.

In \Figref{fig:RGFlowKondoParity} we have used different symbols for the odd and even $M$ results.
The unique feature of the new fixed point is that this odd-even parity effect is lifted and the system flows uniquely
to the even parity fixed point. Note that in the strong coupling regime the odd and even parity sectors are exchanged
due to the formation of the Kondo singlett. In return we can not use the parity to distinguish the ground states.
This is actually a pretty peculiar property, as fermionic operators always appear with an even number of factors in terms
of an Hamiltonian. We have checked that neither a Rashba type spin-orbit coupling nor a BCS singlet or triplet pairing term
in the lead generates such a fixed point, which is obvious as they all conserve fermion parity.
Indeed, in order to break the fermion parity we would need terms consisting of an odd number of creation and annihilation operators.

\begin{figure}[t]
 \includegraphics[width=0.49\textwidth]{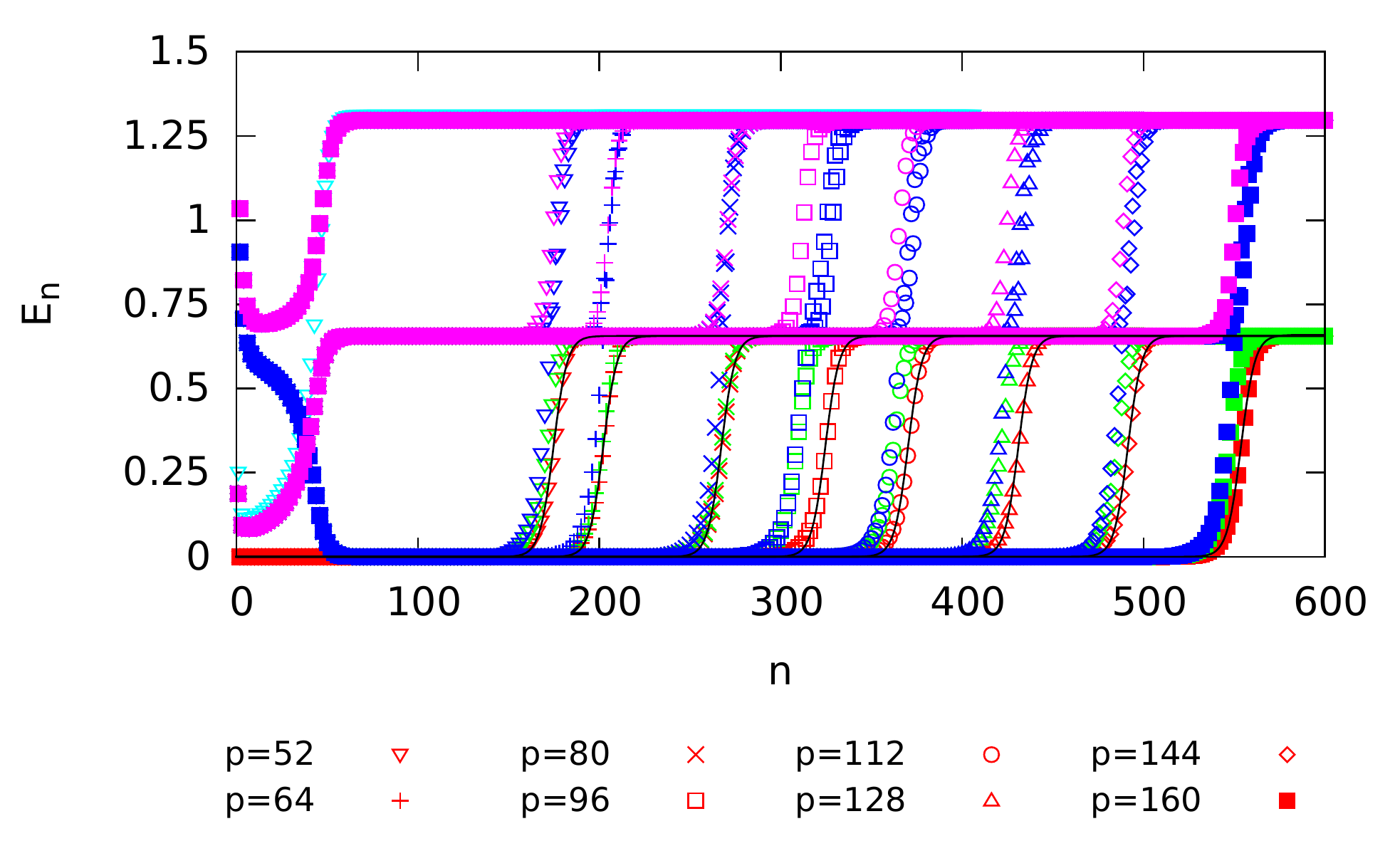} 
 \caption{\label{fig:RGFlowKondoMP} The RG flow of the low energy spectrum for a Kondo model with coupling
 $J=0.25$ vs.\ the NRG step (lead size) $M$,  where the $N$ and $S^z$ conservation is not explicitly enforced,
 $m=400$ states are kept using $\Lambda=2.0$. 
 Calculations are performed using {\tt gmp}/{\tt mpfr} multi precision arithmetic.
 The lines show a fit to the first excited state in the $M$ even sector, which allows to extract the cross over Scale $\Tp$,
 see Fig.~\ref{fig:KondoScaleTp}. }
\end{figure}
While so far the results have been performed in {\tt double} precision, we now also show results for {\tt long double} precision
in \Figref{fig:RGFlowKondoParity}, where the parity lifting fixed point is shifted to a significantly lower energy scale.
At a first sight this looks like an ordinary numerical rounding artefact. However, in that case we expect to obtain noisy
results and not a flow to a new fixed point.  It is important to note that the actual fixed point spectrum is independent
of the the coupling constant $J$, the number of states $m$ kept, provided it is not too small, and the numerical precision.
All these parameter influence the position of the cross--over, not the final energy spectrum, which itself is stable without
further instabilities when continuing the NRG flow. In our opinion these findings suggest that the result has to be
related to a physical effect, otherwise we should not find such a stable result.
The only physical implication of the precision dependent results is given by the mapping of the continuous description
within the Hamiltonian with real or complex numbers numbers onto a discrete world, as the floating point numbers possess
a limited precision only. It is also clear that a truncation due to the discreteness of the representation does not
respect the symmetries of the Hamiltonian and allows breaking all of them. Therefore, the discreteness of the representation
explains how we can break charge and spin conservation leading to a fixed point, where the fermion parity is lifted.
In order to substantiate this interpretation we  investigate the the crossover using the {\tt gmp} multi precision library \cite{gmp}
via the {\tt mpfr} interface \cite{mpfr}. The advantage of the {\tt gmp} based numbers is, that they perform
the numerics with precisely the number of bits requested. In contrast, on the Intel type architecture used for the results
in this work, {\tt double} precision computations performed within the register are often performed with 80 bit resolution, 
and truncated to 64 bit when stored to memory. In addition, fused multiply adds also use higher precision for intermediate
results. Therefore it is hard to actually specify the precise precision of a {\tt double} computation. Similarly it
not obvious, how {\tt long double} arithmetic is actually implemented. The downside of the {\tt gmp} based numerics is,
that it is significantly slower.

In \Figref{fig:RGFlowKondoMP} we show the results obtained using the {\tt gmp} numbers. The results show that the precision
of the numerics sets the cross-over scale, while all calculations flow to the same fixed point. In order to extract the scaling
of the results we fit
\be
	f(M) =  s\, \left( 1 + \tanh\left( \frac{M-\Mp}{w} \right) \right)/2 \label{eq:NRGfit}
\ee
to the first excited state to the results. The fits  shown by lines in \Figref{fig:RGFlowKondoMP} demonstrate that 
\Eqref{eq:NRGfit} is suited to extract the cross over scale.

\begin{figure}[h]
 \includegraphics[width=0.49\textwidth]{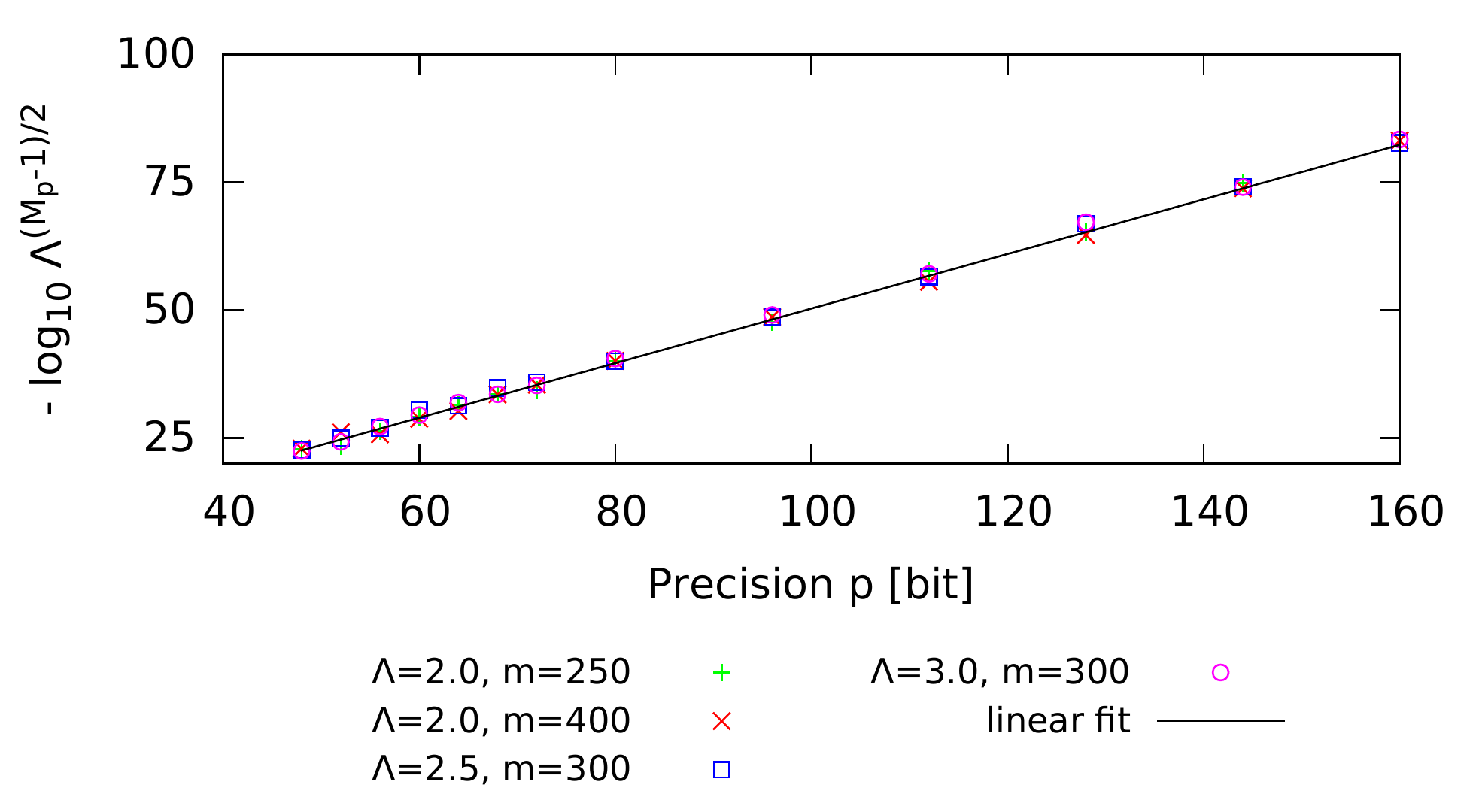} 
 \caption{\label{fig:KondoScaleTp} The cross-over sale $\Tp$ as obtained from fitting \Eqref{eq:NRGfit} as shown in Fig.~\ref{fig:RGFlowKondoMP}
  vs.\ the numeric precision in number of bits $p$ of the mantissa of the floating point variables.
  The pluses correspond to $\lambda=2.0$ and $m=250$, the crosses to  $\lambda=2.0$ and $m=400$, the  squares to $\Lambda=2.5$ and $m=300$,
  and the circles to $\Lambda=3.0$ and $m=300$.}
\end{figure}
In \Figref{fig:KondoScaleTp} we show the cross-over scale obtained from such fits for $\Lambda=2.0$, $2.5$, and $3.0$
and for different number of states $m$ kept. Most strikingly we find a linear behaviour between the precision of
the numerics in bits, i.e.\ $\log_2$ of the actual precision, and the $\log$ of the cross-over scale set by $\Mp$.
The fact that we find the same scaling relation for different $\Lambda$ and $m$ provides strong evidence that our
findings are not just numerical noise. Instead they demonstrate that the results correspond to those of a discrete world.



\section{Summary \& Outlook}

In this work we have shown that a (hypothetical) discreteness at a low energy scale can induce a new fixpoint 
in the RG flow of the Kondo model which displays the unique feature of a broken charge and spin conservation.
Therefore, such a scenario should provide clear signatures for experimental observation.
We would like to note that, although the fixed point found in this work is remarkable, it is based on one specific  scenario
for a discrete world at low energy scales. In the way the NRG works by always blowing up the actual scale our
discreteness is  a relative one, as in the calculations presented here we are not limited by an absolute scale,
just for a process at energy $E$ the world looks discrete on a scale $2^{-p} E$.
One could also interpret the results as describing a world where the commutation laws are not valid at lowest scales.

Yet the stability of the results found in this work suggests, that the symmetry breaking introduced by a discrete energy space,
e.g.\ a finite size universe, should lead to experimentally observable signatures.
In addition we have shown that a discrete world may lead to a fermion parity lifted phase.
From a numerical perspective we can conclude that if someone finds a parity lifted phase in an NRG calculation its
presumably due to a limited precision.
While the use of the standard numerical type is convenient, they have the disadvantage of not being invented to describe 
a discrete world. Due to the exponentials part the discreteness is always relative to  the largest scale, which is here
given by the band width of the system. It will be interesting to study the RG flow using a fixed precision arithmetic
which introduces a physical scale.
   
\section{Acknowledgments}
\begin{acknowledgments}
 I became aware of the appearance of the new fixed point discussed in this work while working on the exercises
 of our computational condensed matter lectures already a few years ago. I have to thank Andreas Poenicke for pushing me forward to
 investigate this phenomenon. 
 Calculations are performed using the Eigen 3 library \cite{eigen3}, the Gnu multi precision library \cite{gmp} via the MPFR interface \cite{mpfr},
 and {\tt g++} from the Gnu compiler collection \cite{gcc}.
\end{acknowledgments}




\begin{thebibliography}{99}
%
\bibitem{Roger:PRL82}     {\sc A. Aspect, P. Grangier}, and {\sc G. Roger},  {\em Experimental realization of Einstein-Podolsky-Rosen-Bohm Gedankenexperiment: A new violation of Bell’s inequalities},  Phys. Rev. Lett. {\bf 49}, 91 (1982).
\bibitem{Gisin:Nature98}  {\sc W. Tittel, J. Brendel, H. Zbinden}, and {\sc Gisin}, {\em Violation of Bell inequalities by photons more than 10 km apart.}, Phys. Rev. Lett. {\bf 81}, 3563 (1998).
\bibitem{Aspect:Nature99} {\sc A. Aspect}, {\em A. Bell’s inequality test: more ideal than ever},  Nature {\bf 398}, 189 (1999).
\bibitem{Zeilinger:PRL98} {\sc G. Weihs, T. Jennewein, C. Simon, H, Weinfurter}, and {\sc A. Zeilinger}, {\em Violation of Bell’s inequality under strict Einstein locality conditions},  Phys. Rev. Lett. {\bf 81}, 5039 (1998).
\bibitem{Wineland:Nature01} {\sc M. A. Rowe, D. Kielpinski, V. Meyer, C. A. Sackett, W. M. Itano, C. Monroe} and {\sc D. J. Wineland}  {\em Experimental violation of a Bell’s inequality with efficient detection},  Nature {\bf 409}, 791 (2001).
%
\bibitem{Bell:P64}        {\sc J. S. Bell}, {\em On the Einstein-Podolsky-Rosen paradox}, Physics {\bf 1}, 195 (1964).
%
\bibitem{Zbinden:Nature08} {\sc Daniel Salart, Augustin Baas, Cyril Branciard, Nicolas Gisin} and {\sc Hugo Zbinden}, {\em Testing the speed of 'spooky action at a distance}, Nature {\bf 454}, 861 (2008).

\bibitem{tHooft:X12} {\sc Gerad 't Hooft}, {\em Relating the quantum mechanics of discrete systems to standard canonical quantum mechanics}, ITP-UU-12/14; SPIN-12/12;  arXiv:1204.4926 (2012).
\bibitem{tHooft:X16} {\sc Gerad 't Hooft}, {\em The Cellular Automaton Interpretation of Quantum Mechanics}, arXiv:1405.1548.
\bibitem{Wheeler:PR55} {\sc J. A. Wheeler}, {\em Geons}, Phys. Rev. {\bf 97}, 511 (1955).
\bibitem{Witten:NPB86} {\sc Edward Witten}, {\em Non-commutative geometry and string field theory}, Nucl. Phys. {\bf B} 268,  253 (1986).
\bibitem{ConnesLott:NPPS89} {\sc A. Connes} and {\sc J. Lott}, {\em Particle models and noncommutative geometry}, Nucl. Phys. Proc. Suppl. B{\bf 18}, 29 (1989).
\bibitem{Hossenfelder:LRR13} {\sc Sabine Hossenfelder}, {\em Minimal Length Scale Scenarios for Quantum Gravity}, Living Rev. Relativity {\bf 16}, 2 (2013).
\bibitem{Itzhak:09}  {\sc  Itzhak Bars} and {\sc John Terning}, {\em Extra Dimensions in Space and Time}, Springer, ISBN 978-0-387-77637-8, 27 (2009).

\bibitem{deHaas:P36}    {\sc W.J. de Haas} and  {\sc G.J. van den Berg}, {\em The electrical resistance of gold and silver at low temperatures}, Physica {\bf 3}, 440 (1936).
\bibitem{Kondo:PTP64}  {\sc J. Kondo},       {\em  Resistance Minimum in Dilute Magnetic Alloys}, Prog. Theor. Phys. {\bf 32}, 37 (1964).
\bibitem{Kondo:SSP69}  {\sc J. Kondo},       {\em Theory of Dilute Magnetic Alloys}, Solid State Phys. {\bf 23}, 183 (1969).
\bibitem{Wilson:RMP75} {\sc K. G. Wilson},   {\em The renormalization group: Critical phenomena and the Kondo problem}, Rev. Mod. Phys. {\bf 47}, 773 (1975).
\bibitem{Wilson:Adv75} {\sc K. G. Wilson},   {\em Renormalization Group Methods},  Adv. in Math. {\bf 16}, 170 (1975).
\bibitem{Hewson:93} {\sc A. C. Hewson} Hewson,{\em The Kondo Problem to Heavy Fermions}, Cambridge University Press, Cambridge, ISBN-13: 9780521599474(1997).
\bibitem{CostiPruschke:RMP08} {\sc T. Costi} and {\sc T. Pruschke}, {\em The numerical renormalization group method for quantum impurity systems}, Rev. Mod. Phys. 80, 395 (2008).
\bibitem{Andrei:PRL80} {\sc N. Andrei},      {\em Diagonalization of the Kondo Hamiltonian}, Phys. Rev. Lett. {\bf 45}, 379 (1980).
\bibitem{Wiegman:JETP80} {\sc  P.B. Vigman}, {\em Exact solution of s-d exchange model at T = 0}, Sov. Phys. JETP Lett. {\bf 3}, 392 (1980).
%
\bibitem{GoldhaberGordon:PRL98} {\sc D. Goldhaber-Gordon, J. Gores, M. A. Kastner, H. Shtrikman, D. Mahalu}, and {\sc U. Meirav}, {\em }, Phys. Rev. Lett. {\bf 81}, 5225 (1998).
\bibitem{MetznerVollhardt:PRL89} {\sc W. Metzner} and {\sc D. Vollhardt},  Phys. Rev. Lett. {\bf 62}, 324 (1989).
\bibitem{Georges:RMP96} {\sc A. Georges, G. Kotliar, W. Krauth}, and {\sc M. J. Rozenberg}, Rev. Mod. Phys. {\bf 68}, 13 (1996).
%
%
\bibitem{IEEE754} {\sc ANSI/IEEE STD 754-2008}, {\em 754-2008 -- IEEE Standard for Floating-Point Arithmetic}, E-ISBN 978-0-7381-5752-8, DOI 10.1109/IEEESTD.2008.4610935 (2008).
%
\bibitem{gmp}  {\sc Torbj\"orn Granlund} and {the GMP development team},{\em GNU MP: {T}he {GNU} {M}ultiple {P}recision {A}rithmetic {L}ibrary}, {\tt http://gmplib.org/} (2012).
\bibitem{mpfr} {\sc L. Fousse, G. Hanrot, V. Lef\`{e}vre, P. P{\'e}lissier}, and {\sc P. Zimmermann}, {\em MPFR: A multiple-precision binary floating-point library with correct rounding}, ACM TOMS {\bf 33}, {\tt  http://www.mpfr.org.} (2007).
\bibitem{eigen3} {G\sc a\"{e}l Guennebaud and Beno\^{i}t Jacob} et. al., {\em Eigen v3}, {\tt http://eigen.tuxfamily.org} (2010).
\bibitem{gcc} {\sc R. M. Stallman} et al., {\em Using the GNU Compiler Collection"}, Free Software Foundation, Boston, {\tt https://gcc.gnu.org} (2012).
%
\end{thebibliography}
\end{document}